\documentclass[prl,twocolumn,floatfix,showpacs,superscriptaddress,amsmath,amssymb]{revtex4}
\usepackage{color}
\usepackage{units}
\usepackage{verbatim}
\usepackage{graphicx}
\definecolor{ngreen}{rgb}{0.2,0.6,0.2}%272

\newcommand{\blk}{\color{black}}

\newcommand{\beq}{\begin{equation}}
\newcommand{\eeq}{\end{equation}}
\newcommand{\bqa}{\begin{eqnarray}}
\newcommand{\eqa}{\end{eqnarray}}

\newcommand{\erf}[1]{Eq.~(\ref{#1})}

\newcommand{\ea}{{\it et al.}}

\newcommand{\dg}{^\dagger}

\newcommand{\smallfrac}[2]{\mbox{$\frac{#1}{#2}$}}
\newcommand{\half}{\smallfrac{1}{2}}

\newcommand{\bra}[1]{\langle{#1}|}
\newcommand{\ket}[1]{|{#1}\rangle}

\newcommand{\op}[2]{\hat \tau_{#1}^{#2}}

\newcommand{\sq}[1]{\left[ {#1} \right]}

\newcommand{\ro}[1]{\left( {#1} \right)}
\newcommand{\an}[1]{\left\langle{#1}\right\rangle}

\newcommand{\tr}[1]{{\rm Tr}\sq{ {#1} }}

\newcommand{\real}{\Re}

\renewcommand{\d}{^{\rm Y}}
\newcommand{\p}{^{\rm S}}
\newcommand{\ob}{^{\rm O}}

\newcommand{\q}{^{\rm Q}}

\newcommand{\s}[1]{\hat \sigma_{#1}}

\renewcommand{\Re}{{\rm Re}}

\renewcommand{\section}[1]{{\em #1}.---}

\begin{document}

\title{Are dynamical quantum jumps  detector-dependent?}

\author{Howard M. Wiseman}
\affiliation{Centre for
Quantum Computation and Communication Technology (Australian Research Council), \\ Centre for Quantum Dynamics, Griffith University, Brisbane, Queensland 4111, Australia}%\email{h.wiseman@griffith.edu.au}
\author{Jay M. Gambetta}
\affiliation{IBM T. J. Watson Research Center, Yorktown Heights, New York 10598, USA}
%\email{jay.gambetta@us.ibm.com}

\begin{abstract}
Dynamical quantum jumps were initially conceived by Bohr  as objective events associated with the {\em emission} of a light quantum by an atom. Since the early 1990s they have come to be understood as being associated rather with the {\em detection} of a photon by  a measurement device, and that different detection schemes result in different types of jumps (or diffusion). 
Here we  propose experimental tests to rigorously prove the  detector-dependence of the stochastic evolution of an individual atom. 
The tests involve no special preparation of the atom or field, and the required efficiency can be as low as $\eta \approx 58\%$. 
\end{abstract}


\maketitle

Quantum jumps---the discontinuous change in the state of a microscopic system (such as an atom) at random times---were the first form of quantum {\em dynamics} to be introduced \cite{Boh13,Ein17}. 
This was in the 1910s,  long before the notion of entanglement, and its puzzles such as the Einstein-Podolsky-Rosen (EPR) paradox \cite{EPR35}, had been  introduced, and before the special role of measurement in quantum mechanics had been elucidated \cite{Von32}. Thus these 
jumps were conceived of as objective dynamical events, linked to an equally objective photon emission event.  

The modern concept of quantum jumps  in atomic systems  is based on ``quantum trajectory'' theory \cite{Car93b}, introduced independently in  Refs.~\cite{DalCasMol92,GarParZol92,Bar93}.  This theory comprises stochastic evolution equations for the microscopic system state conditioned on the results of monitoring the bath to which it is weakly coupled.  
These are also known as ``unravellings''   \cite{Car93b} of the system's master equation (ME), as the ensemble average  of the quantum trajectories of the (ideally pure) state of an individual system replicates the mixture-inducing evolution of the ME.  
This theory  
has been applied also to solid-state qubits and other quantum systems \cite{WisMil10,KorPRB99,Gambetta2008}. In the atomic case, a photodetection event causes the state of the distant atom to  jump  because of entanglement between the bath (the electromagnetic field) and the atom.  That is, the quantum jumps are 
{\em detector-dependent};  in the absence of a measurement there would be no jumps.  This is in marked contrast to the  objective  jumps of Bohr and Einstein,  linked to a supposed emission event. 

One could well be tempted to ask 
what difference does it make if we say a jump is caused 
by the emission of a photon rather than saying it is caused by the detection 
of a photon? 
If photon detection were the only way to measure the emitted quantum 
field then the answer %in both cases 
would indeed be: no real difference at all. But there are many 
(in fact infinitely many) other ways to measure the emitted field. For instance one can  
interfere it with a local oscillator (LO)---that is, another optical field in a coherent state---prior 
to detection \cite{Car93b}. 

In quantum optical detection theory  
radically different stochastic dynamics for the atomic state occur depending on the detection 
scheme used by a distant observer \cite{Bar93,Car93b,WisMil10}. Perfect ``direct'' (no LO) photon detection gives rise to the quantum jump model of Bohr and Einstein. Perfect heterodyne detection (with a strong, far-detuned LO), by contrast, gives rise to a particular 
type of pure-state quantum {\em diffusion} \cite{WisMil93c}, originally proposed as an 
objective (i.e. detector-independent) model 
of ``quantum state diffusion'' (QSD) \cite{GisPer92b}.  It is not that either the quantum jump model or 
the QSD model is wrong. Rather it is the fact that both are 
valid unravellings of the same ME that says that neither of them are {\em objective pure-state dynamic models} (OPDM); 
according to quantum optical measurement theory, no such model can exist. 

In this Letter we drop the assumption that quantum optical measurement theory is correct in order 
to ask whether, and how, one could try to rule out these OPDMs {\em experimentally}, 
Moreover, we do not want to rule out just the quantum jump model and the QSD 
model as detector-independent models; we want experiments that could rule out {\em all} OPDMs. 
This would prove that quantum jumps (or diffusion) is measurement-dependent.

We propose experimental tests on a resonantly driven two-level atom that could rule out all OPDMs for the atom, 
and that do not require any special preparation of the atom or field. The key to these tests is the ability to implement two different ways of monitoring the radiated field, giving rise to two different sorts of stochastic evolution. This is an instance of the EPR phenomenon, also known as ``steering'' \cite{SchPCP35}, when understood sufficiently generally \cite{WisJonDoh07}. Specifically, under the assumption that an objective state exists (obeying some OPDM), we derive an {\em EPR-steering inequality} \cite{CavJonWisRei09} that could be violated experimentally \cite{SJWP10,Smi12}. With two particular monitoring schemes we propose, this could be done for efficiencies as low as $0.58$, which is the main result of this Letter.  

This Letter is organized as follows. First we briefly explain how EPR-steering can be demonstrated experimentally. 
Then we present the model quantum system for our investigation: resonance fluorescence of a strongly driven two-level atom. Next we present two very different monitoring schemes: a spectrally resolved photon counting (jump) scheme; 
and a homodyne (diffusion) scheme. 
We then show that an EPR-steering inequality suitable for these continuous-in-time measurements can be violated for an efficiency $\eta>0.58$. We also consider the  
option of using two different homodyne schemes (X and Y). Although the minimum sufficient efficiency is somewhat higher 
in this case ($\eta>0.73$), this test would probably be more practical. Finally we derive a {\em necessary} efficiency condition $\eta > 1/2$
 which pertains even if one could implement the whole class of diffusive unravellings.  
 
\section{EPR-steering} 
In the original EPR paradox \cite{EPR35}, the ability of one party (Alice) to measure different 
observables---position and  momentum---on her half of an entangled state allows her 
to collapse the state of the other half (Bob's) into different and 
incompatible states---position and momentum eigenstates---at will. That is, the essence of the phenomenon
is that Alice can disprove the hypothesis that Bob's system has an objective quantum state (i.e. one 
existing independent of her measurement choice) \cite{WisJonDoh07}. It is this general 
phenomenon of EPR-steering  that lies behind our tests. 

In our case Bob has a two-level system, or qubit. For any qubit state $\rho$ it is easy to verify that 
\beq
f_1(\rho) + f_2(\rho) \leq 1,
\label{ineq-1}
 \eeq 
with $
f_1(\rho) \equiv \ro{ \tr{\s{x}\rho} }^2$, $ f_2(\rho) \equiv
 \ro{ \tr{\s{y}\rho} }^2+ \ro{ \tr{\s{z}\rho} }^2$. 
Now under the objective quantum state assumption, Bob's system is in some pure state 
$\rho\ob_\psi = \ket{\psi}\bra{\psi}$. This state may be unknown to him, and chosen at 
random from an ensemble $\{\tilde\rho\ob_\psi \}$.  Here we are using unnormalized 
states $\tilde\rho\ob_\psi = p\ob_\psi \rho\ob_\psi$, where 
$p\ob_\psi d\mu(\psi)$ is the probability distribution over the pure states, with 
$d\mu(\psi)$ being a standard measure.  Now \erf{ineq-1} applies 
for all $\rho$, so  
\beq \label{ineq-2}
 {\rm E}[f_1(\rho\ob)] + {\rm E}[f_2(\rho\ob)] \leq 1,
\eeq
where $ {\rm E}[f(\rho\ob)]  \equiv  \int d\mu(\psi)  p\ob_\psi f(\rho\ob_\psi)$.

Say now that Alice can choose between two different measurements, called S and Y (for reasons that will become apparent later). 
Under the objective state assumption, Alice's measurement can do nothing except provide some information
about {\em which} state $\rho\ob_\psi$ pertains to Bob's system. That means that 
the ensemble of possible states for Bob's system, conditioned on the result of Alice's measurement S,
 is a {\em coarse-graining} of Bob's objective pure state ensemble \cite{JonWisDoh07}. Since $f_1$ is 
 convex on the state-space of $\rho$, it follows that ${\rm E}[f_1(\rho\p)] \leq {\rm E}[f_1(\rho\ob)]$ \cite{CavJonWisRei09}. 
 The same remarks hold for the Y-ensemble and $f_2$, so ${\rm E}[f_2(\rho\d)] \leq {\rm E}[f_2(\rho\ob)]$. Thus from \erf{ineq-2} we  have 
\beq \label{ineq-3}
S^{\rm S,Y} \equiv {\rm E}[f_1(\rho\p)] + {\rm E}[f_2(\rho\d)] \leq 1.
\eeq

The two terms in this EPR-steering inequality can be measured experimentally by Alice choosing between 
measurements S and Y, and by Bob correlating the results of measurements on his qubit  
with the results Alice reports. If \erf{ineq-3} is violated 
 then the experiment disproves the hypothesis that Bob has an objective pure state. 
This is possible only if (as in the EPR paradox) Bob's system is entangled with Alice's prior to her measurement.  
Further details on the above argument, in particular how it works when each of Alice's measurements is 
a continuous-in-time readout of a quantum field, are given below.

\section{Example quantum optical system} Let Bob's quantum system be a single two-level atom, driven by a resonant laser field. This is described by the resonance fluorescence 
ME; in the usual interaction frame \cite{WisMil10}, which removes the atomic transition frequency $\omega_0$, this is:
\beq \label{me-1}
\dot\rho = {\cal L}\rho \equiv -i[\hat H_\Omega,\rho]   + \gamma {\cal D}[\s{-}]\rho.
\eeq
Here $\hat H_\Omega = (\Omega/2)\s{x}$ is the Hamiltonian describing resonant driving, $\s{-}=(\s{x}-i\s{y})/2$ is the atomic lowering operator, and ${\cal D}[\hat c]\rho \equiv \hat c\rho \hat c\dg - \half(\hat c\dg \hat c \rho + \rho \hat c\dg \hat c)$ as usual.  Now move into the $\Omega$-rotating frame with respect to the Hamiltonian 
$\hat H_\Omega$. This effects the transformation 
\beq \label{eq:rf}
\s{-} \to  \frac{1}{2}\left( \op-+ e^{-i\Omega (t-t_0)}  + \s{x}  - \op+- e^{+i\Omega (t-t_0)} \right),
\eeq
 where  $\op{+}{-} = (\op{-}{+})\dg \equiv \ket{+}\bra{-}$, and $\s{x}\ket{\pm}=\pm\ket{\pm}$.
 Now if $\Omega \gg \gamma$ then in this rotating frame we can make the secular approximation, dropping terms in the ME rotating at frequency $\Omega$ or $2\Omega$. This gives the approximate ME   \beq
\dot\rho = \bar{\cal L}\rho  \equiv \frac{\gamma}{4}\left( {\cal D}[\op-+] + {\cal D}[\s{x}] + {\cal D}[\op+-]\right) \rho . \label{me-2}
\eeq 
The simplicity of this equation allows us to obtain the semi-analytical results for EPR-steering below. Some experimental considerations are discussed in Ref.~\cite{epaps}.

\section{Spectral Adaptive Interferometric Detection (SAID)}
The three irreversible terms in \erf{me-2} correspond to the three spectral peaks in resonance fluorescence, at frequencies $\omega_0-\Omega$, $\omega_0$, and $\omega_0+\Omega$ respectively.  
If Alice uses a spectrally-resolving detection technique in the limit of large $\Omega$, the atom will undergo three types of jump \cite{WisToo99}. 
The jump operator from the first term ($\op-+$) collapses the atom into the $\ket{-}$ state; that from the last ($\op+-$) into the $\ket{+}$ state. These are ``good'' jumps in terms of making the system state pure. The jump operator from the middle term ($\s{x}$) however does not change the purity of the state. This makes it a ``bad'' jump because, if the efficiency is less than unity,  the purity of the system state decays monotonically following a jump, and can be restored only when it next jumps.  
However these ``bad'' jumps can be made ``good'' if Alice adds a weak LO to the fluorescence before detection, resonant with the atom, so that the source field is proportional not to $\s{-}$, but to $\s{-} \pm 1/2$. Here the two cases correspond to opposite signs for the LO. 
Then the jump operator from the middle term becomes proportional to $\s{x}\pm 1$, which equals $2\hat\pi_\pm$, 
where $\hat\pi_\pm = \ket{\pm}\bra{\pm}$. 

The optimal scheme, for maximizing the purity of the state, is to choose the LO phase {\em adaptively}, using real time feedback \cite{WisMil10}. Specifically, Alice should choose the $\pm$ case when the preceding  jump put the atom into the $\ket{\pm}$ state. This choice maximizes the rate of jumps, which is optimal because every jump repurifies the atom. 
Following a jump into state $\ket{+}$, the quantum trajectory theory appropriate to an efficiency $\eta_{\rm S}\leq 1$ \cite{WisMil10} predicts that the unnormalized state will evolve according to 
\beq \label{nospectraljumps}
\dot{\tilde\rho}\p  =  \bar{\cal L}\tilde \rho\p - \frac{\gamma\eta_{\rm S}}{4}
\ro{{\cal J}[\op-+] +    4{\cal J}[\hat\pi_+] + {\cal J}[\op+-]} \tilde\rho\p,
\eeq
where the `S' superscript means the state is conditioned on this SAID scheme and  ${\cal J}[\hat c]\rho \equiv \hat c\rho \hat c\dg$ as usual. This state is unnormalized, with 
the decaying norm equal to the probability of lasting so long without another jump. 

\begin{figure}\begin{center}
\includegraphics[width=0.85\linewidth]{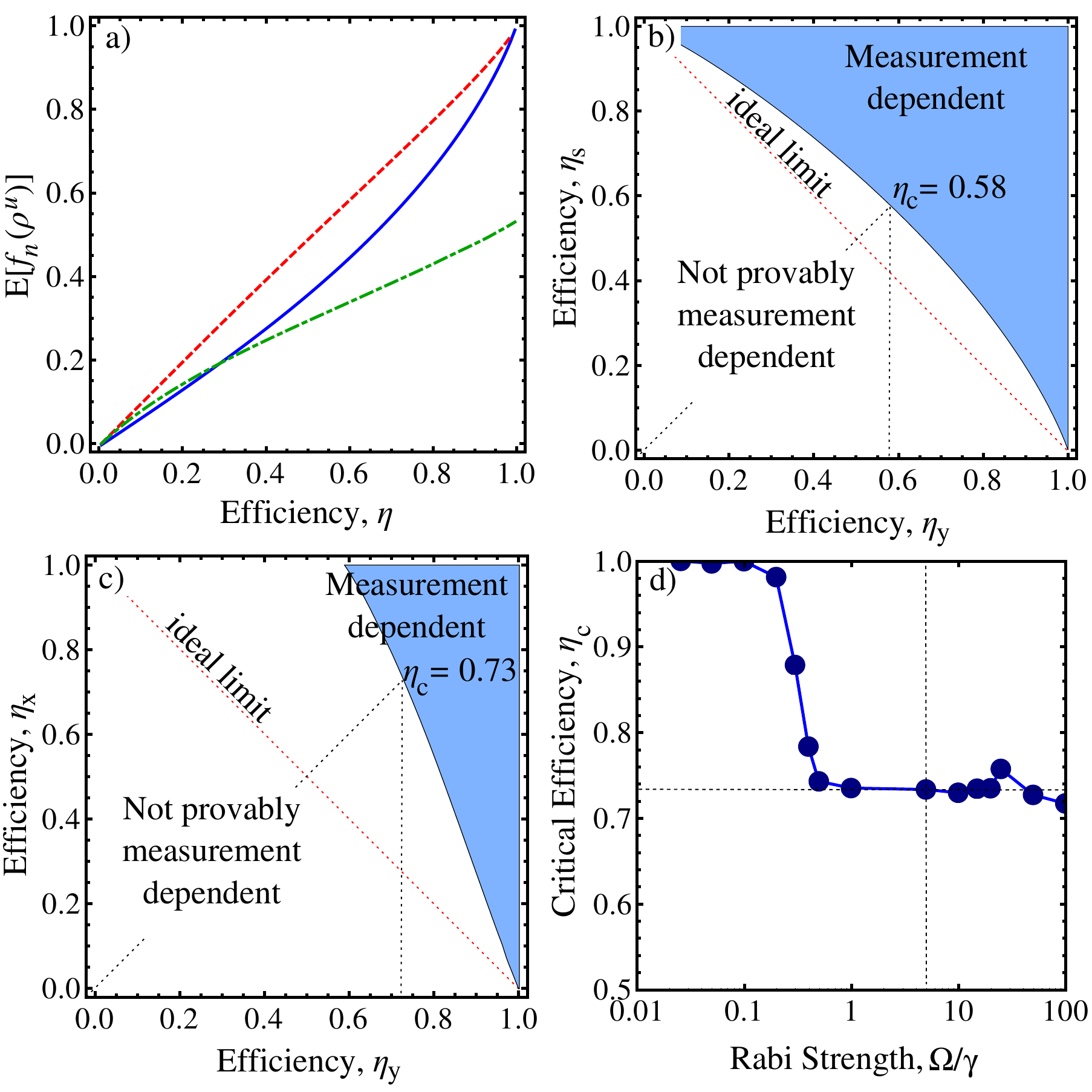}\end{center}
\vspace{-3ex} \caption{(Color online) a) As a function of efficiency $\eta$, we plot: (red dashed) ${\rm E}[(\an{\s{x}}\p)^2]$ for the SAID unravelling, large $\Omega$ limit;  (solid blue) ${\rm E}[(\an{\s{y}}\d)^2 + (\an{\s{z}}\d)^2]$ for the Y-homodyne unravelling, in the large $\Omega$ limit; (green dot-dashed) ${\rm E}[(\an{\s{x}}\p)^2]$ for the X-homodyne unravelling from simulations with $\Omega=5\gamma$. b) $S^{\rm S,Y}(\eta_{\rm Y},\eta_{\rm S})$  is plotted with the blue shaded regime ($S>1$) representing when the experiment would rule out all theories of objective atomic state reduction. c) is the same as b) with  SAID replaced by X-homodyne detection and $\Omega=5\gamma$ for both unravellings. d) The efficiency required for X- and Y-homodyne to have $S>1$, as a function of $\Omega/\gamma$. 
For more details see text.}
\label{fig:Measurements}\label{fig:saidhomoy}\end{figure}

It is easy to verify that the solution of \erf{nospectraljumps} is a mixture of $x$-eigenstates, and that each 
jump repurifies it to $x=\pm 1$. Thus the stationary ensemble $\{\tilde\rho\p_x\}$ of conditional states under the SAID scheme %(indicated by a `s' superscript) 
is indexed by a single real parameter, $x=\an{\s{x}}$. We can quantify how well the scheme maintains the state purity by ${\rm E}[(\an{\s{x}}\p)^2]$, the ensemble average of the square of $x$. We show in Ref.~\cite{epaps} how to  
calculate this analytically. 
The result is shown as a function of $\eta_{\rm S}$, in Fig.~\ref{fig:saidhomoy} a) as a red dashed line. As expected, it is monotonically increasing with $\eta_{\rm S}$ and attains unity as $\eta_{\rm S}\to 1$.
 
\section{Y-Homodyne Detection} As our second unravelling of the ME (\ref{me-1}) we consider homodyne detection. 
The quantum trajectories for the {\em normalized} conditioned state under this monitoring are obtained by adding the following 
 (zero-mean)  term to the ME (\ref{me-1}) \cite{WisMil10} 
\beq \label{dyne}
d\rho\q = \sqrt{\gamma}{\cal H}[ dZ^*(t) \s{-}]\rho\q .
\eeq 
Here ${\cal H}[\hat c] \equiv \hat c\rho + \rho \hat c - {\rm Tr}[(\hat c+ \hat c\dg)\rho]$, and 
$Z(t)$ is a complex random Wiener increment (related to the noise in the photocurrent), normalized so that $\an{ |dZ(t)|^2}= \eta_{\rm Q}dt$. The phase $\phi$ of the LO appears in $\an{dZ(t)^2} = e^{2i\phi} \eta_{\rm Q} dt$. This allows different quadratures (`Q') to be monitored, and the Y-quadrature corresponds to $\phi = \pi/2$.  

For homodyne detection to maintain a high-purity conditioned atomic state in the $\Omega \gg \gamma$ limit requires the effective bandwidth \cite{WisMil10} of the detector to be much greater than $\Omega$, as the spectrum of the homodyne photocurrent will have a signal at zero frequency and at $\pm \Omega$~\cite{WisMil93c}.
These correspond to the three terms in \erf{me-2}, and in that rotating frame the additional terms describing conditioning on Y-homodyne detection are 
\beq
d\rho\d = \sqrt{\frac{\eta_{\rm Y}\gamma}{4}}{\cal H}[ idV_\Omega \op-+  + idW_x\s{x} - i dV_\Omega^* \op+- ] \rho\d . \label{me-2}
\eeq
Here $dV_\Omega$ is an  {\em irreducibly complex} Wiener increment satisfying $\an{|dV_\Omega|^2}=dt$ but $\an{dV_\Omega^2}=0$, independent of the {\em real} increment $dW_x$. The minus sign before the last term comes from the corresponding minus sign in \erf{eq:rf}.

It is not difficult to show that the long-time solution of this Y-homodyne quantum trajectory is confined to the $x=0$ plane of the Bloch-sphere, and so is parametrized by two numbers, $y=\an{\s{y}}$ and $z=\an{\s{z}}$. That is, the ensemble $\{\tilde\rho\d_{(y,z)}\}$ of conditioned states 
in this case is as different as possible from the ensemble $\{\tilde\rho\p_x\}$ in the SAID monitoring case, which  is  confined to the $x$-axis. As above, we can quantify how well this monitoring scheme maintains the state purity by the long-time expectation for the square of the conditional average of the Bloch vector length, ${\rm E}[(\an{\s{y}}\d)^2 + (\an{\s{z}}\d)^2]$. As we show in Ref.~\cite{epaps}, this can easily 
be calculated to arbitrary numerical precision.  The result is shown in Fig.~\ref{fig:saidhomoy} a) as the solid blue line. It is similar to that for SAID, but is smaller (the conditioned states tend to be less pure). 

\section{EPR-Steering criterion} 
Assume now, as imagined in OPDMs, that the atom has some `true' pure state 
$\rho\ob_\psi = \ket{\psi}\bra{\psi}$ at every instance of time. The index 
$\psi$ that pertains at any particular time changes stochastically in some manner 
defined by the model.   In the standard quantum jump model  $\psi$ 
would  evolve deterministically and smoothly apart from the occasional jump, while in the QSD model, 
for example, $\psi$ would evolve non-deterministically and non-smoothly 
at all times.  
Once the system has reached equilibrium it will be described by a stationary ensemble $\{\tilde\rho\ob_\psi  \}$ 
as described earlier. 

Now under the above OPDM assumption, the stationary ensemble $\{\tilde\rho\p_x \}$ 
 realized by Alice's SAID scheme must be merely a coarse-graining of 
$\{\tilde\rho\ob_\psi  \}$. That is, there must exist a conditional probability distribution $\wp(x|\psi)$ 
such that $\forall x, \tilde\rho\p_x = \int d\mu(\psi) \wp(x|\psi) \tilde\rho\ob_\psi$.
The same statements hold for the Y-homodyne ensemble $\{\tilde\rho\d_{(y,z)}\}$, {\em mutatis mutandis}. Thus 
from the arguments rehearsed earlier, we can derive the EPR-steering inequality \erf{ineq-3}, where the value of $S^{\rm S,Y}$  depends only upon the efficiencies. 

Experimentally, the quantity $S^{\rm S,Y}(\eta_{\rm S},\eta_{\rm Y})$ could be measured as follows. In any given run, Alice randomly chooses to implement either the SAID or the Y-homodyne monitoring of the atomic fluorescence. After some randomly chosen time $T \gg \gamma^{-1}$, Bob halts the experiment (so that no further light reaches Alice), and immediately reads out the state of the atom. More particularly, he randomly measures either $\s{x}$, $\s{y}$ or $\s{z}$.  Alice reveals to Bob  which monitoring scheme she performed, and the parameters ($x$ in case 1, and $y, z$ in  case 2) of her conditioned state for the atom at time $T$, which she can determine at her leisure from her measurement record using the above quantum trajectory theory. Bob then stores the results of his measurements in different `bins' for different values of $x$ (in case 1) or $y, z$ (in case 2). He can then calculate, for example, the first term in \erf{ineq-3} by first determining the average of his measurement of $\s{x}$ in each bin, then squaring it, then averaging over all bins, weighted by the number of entries in each bin. 

Note that Bob does not have to put any trust in the state parameters that Alice reports; they are merely labels for his bins. Nevertheless, for sufficiently fine binning, if Alice really can implement the above detection schemes, we would expect the result of Bob's averaging procedure to agree with the ensemble averages ${\rm E}[(\an{\s{x}}\p)^2]$ and ${\rm E}[(\an{\s{y}}\d)^2 + (\an{\s{z}}\d)^2]$ computed above. If the results {\em violate} the inequality (\ref{ineq-3}), then the experiment would have proven that 
the initial assumption, that there exists an OPDM for the atom, must be wrong. In Fig.~\ref{fig:saidhomoy} b) we plot $S^{\rm S,Y}(\eta_{\rm Y},\eta_{\rm S})$, and see that the EPR-steering inequality (\ref{ineq-3}) is violated in the blue shaded regime. 
In particular, if $\eta_{\rm Y} = \eta_{\rm S} = \eta$, one could prove that quantum jumps are  detector-dependent with 
 $\eta\gtrsim 0.58$. This is the main result of this paper. 
 
Achieving  $\eta\gtrsim 0.58$ is still very challenging experimentally. One might hope that choosing a different pair of unravellings, or using a better EPR-steering inequality, could substantially lower this requirement. This is not the case, as we now show.  If $\eta_{\rm Y} + \eta_{\rm S} \leq 1$ then it would be possible for a hypothetical observer to implement {\em simultaneously} the SAID and the Y-homodyne monitorings, giving rise to a doubly-conditioned ensemble $\{\tilde\rho^{\rm S+Y}_{(x,y,z)}\}$. This ensemble has exactly the right properties to be an OPDM ensemble  $\{\tilde\rho\ob_\psi\}$, since coarse graining by ignoring the Y-conditioning would give $\{\tilde\rho\p_x\}$ and {\em vice versa}. Thus  to disprove  all OPDMs we clearly need $\eta_{\rm Y} + \eta_{\rm S} > 1$. That is, the necessary condition is $\eta > 0.5$, 
 scarcely less onerous than the sufficient condition of  $\eta\gtrsim 0.58$. 
 
\section{X- and Y-Homodyne Detection} In practice, photon counters (at least fast ones, as required here) are less efficient than the photoreceivers used for homodyne detection \cite{WisMil10}. The SAID scheme has the additional challenges of spectral resolution, and feedback much faster than the atomic lifetime $\gamma^{-1}$. Also, the above analysis is valid only in the limit $\Omega / \gamma \to \infty$. For these reasons we now consider replacing the SAID scheme by X-homodyne detection, and we keep $\Omega/\gamma$ finite. Although the X-homodyne scheme does not confine the conditioned system state to the $x$-axis (as in the SAID unravelling) it does tend to do so \cite{WisMil93c,WisMil10}. Thus we expect that for a high enough efficiency we could violate the same EPR-steering inequality ({\ref{ineq-3}),  but with `S' replaced by `X'.
The quantum trajectories for the X- and Y-homodyne schemes are obtained by adding \erf{dyne} to \erf{me-1}, with the LO phases $\phi = 0$ and $\pi/2$ respectively.  Numerical simulations of $S^{\rm Y, X}(\eta_{\rm Y},\eta_{\rm X})$ are shown in Fig.~\ref{fig:Measurements} c) for $\Omega = 5\gamma$ and yield a violation (blue shaded regime) for $\eta \gtrsim 0.73$.  Furthermore this is the critical efficiency for all $\Omega >\gamma$, as shown in Fig.~\ref{fig:Measurements} d).

\section{General diffusive unravellings} With two homodyne schemes of efficiency $\eta$ the same reasoning as earlier implies that a necessary condition for proving the  subjectivity of quantum unravellings is $\eta>1/2$. Now if one can measure the X and Y quadratures then typically one can measure any quadrature, and one might think that being able to choose between say $N$ different quadratures would make it easier to demonstrate EPR-steering, as the obvious necessary condition would be $\eta  > 1/N$. In fact,  $\eta>1/2$ is still necessary,  as shown in Ref.~\cite{epaps}. 

To conclude, the original notion of quantum jumps, from almost 100 years ago, is that atoms radiate via an objective stochastic process. Although quantum trajectory theory developed in the early 1990s asserts that there can be no detector-independent pure-state 
atomic dynamics---regardless of the nature of jumps (or diffusion) proposed---
no experiment has ever been done to rigorously test this assertion. 
With the experimental tests we have proposed here,
 it should be possible to rule out all such objective models, proving finally that dynamical quantum jumps are indeed detector-dependent. 

HMW was supported by the ARC Centre of Excellence grant CE110001027. 

\begin{widetext}

\subsection{\large Supplemental Material for {\em Are dynamical quantum jumps  detector-dependent?}}

\subsection*{\large Experimental Considerations}

For all of the monitoring schemes we consider, we require that a large majority of the fluorescence of the atom 
can be detected by Alice, in the form of a beam (or at least a small number of beams), so that a local oscillator 
may be added easily prior to detection. This could be realized by a trapped ion surrounded by large-aperture lenses
or by an atom coupled (with strength $g$) to a cavity with large damping rate $\kappa$, such that the atomic damping via the cavity output beam is the dominant part of the total spontaneous emission rate $\gamma = \gamma_{\rm free} + g^2/\kappa$ \cite{MabDoh02}. In terms of Bob's tomographic measurement of the atom, this can be done with high efficiency 
for the case of a trapped ion, using the technique of electron shelving \cite{Ber86}. However it is not necessary 
for this measurement to be high efficiency, as Bob can always compensate for low efficiency or otherwise imperfect 
measurements when he takes his ensemble average \cite{Smi12}.

\subsection*{\large Spectral Adaptive Interferometric Detection (SAID)}

Without loss of generality, consider the case where the system state jumped into state $\ket{+}$ at some time $t_j$, and that this information is used to set the LO as above, until the next jump at time $t_{j+1}$. 
 Then, using the quantum trajectory theory appropriate to an efficiency $\eta_{\rm S}\leq 1$ \cite{WisMil10}, 
 the system state for times $t=t_j+\tau < t_{j+1}$  evolves according to 
\beq \label{nospectraljumps}
\dot{\tilde\rho}\p(\tau) =  \bar{\cal L}\tilde \rho\p - \frac{\gamma\eta_{\rm S}}{4}
\ro{{\cal J}[\op-+] +    4{\cal J}[\hat\pi_+] + {\cal J}[\op+-]} \tilde\rho\p.
\eeq
Here the `s' superscript means the state is conditioned on this SAID scheme, 
and  ${\cal J}[\hat c]\rho \equiv \hat c\rho \hat c\dg$ as usual. Because this is a linear equation, the solution 
is easy to find analytically. 
It is unnormalized (indicated by the tilde), with the norm at time $t_j+\tau$ being the probability for no detections in the interval $[t_j,t_j+\tau)$. In other words, the probability density for the next jump happening at time $t_{j+1} = t_j+\tau$ is $p\p(\tau)d\tau = - {\rm Tr}[\dot{\tilde\rho}\p ] d\tau$. Until that point the conditioned state is $\rho\p(\tau) = {\tilde\rho}\p(\tau) / {\rm Tr}[{\tilde\rho}\p(\tau) ]$, which for the appropriate initial condition ${\tilde\rho}\p(0) = \ket{+}\bra{+}$ is a mixture of $x$-eigenstates. At time $t_{j+1}$, the system either resets to $\ket{+}$ so that the above analysis repeats, or jumps to $\ket{-}$ and the above analysis repeats with $\ket{+}$ and $\ket{-}$ swapped. 

Since the ensemble of possible conditional states under the SAID scheme is indexed by a single parameter, $x=\an{\s{x}}$, 
we can characterize how good the scheme is at maintaining a pure state by calculating the time-average of the square of $x$. 
From the above, we have
\beq \label{avxsq}
{\rm E}[(\an{\s{x}}\p)^2] = - \int_0^\infty  ({\rm Tr}[{\rho}\p(\tau) \s{x}])^2  {\rm Tr}[\dot{\tilde\rho}\p(\tau)] d\tau
\eeq
where E stands for expectation value. This integral can also be done analytically, but the expression is too complicated to be usefully displayed.

\subsection*{\large Y-Homodyne Detection}

The stochastic conditional master equation 
\beq
d\rho\d = \frac{\gamma}{4}\left( {\cal D}[\op-+] + {\cal D}[\s{x}] + {\cal D}[\op+-]\right) \rho\d dt +  \sqrt{\frac{\eta_{\rm Y}\gamma}{4}}{\cal H}[ idV_\Omega \op-+  + idW_x\s{x} - i dV_\Omega^* \op+- ] \rho\d . \label{sme-2}
\eeq
has a solution confined to the $y$--$z$ plane. It is convenient to change to a polar co-ordinate system with angle $\theta$ and modulus-squared $\beta=\an{\s{y}}^2+\an{\s{z}}^2$, as there is a simple stochastic differential equation (SDE) for $\beta$: 
\beq
d\beta = \gamma A(\beta)dt + \sqrt{\gamma B(\beta)}\,dW_\beta, \label{betaSDE}
\eeq
 where $A(\beta) = -3\beta/2 + \eta_{\rm Y}(1+\beta^2/2)$, $B(\beta) = 2\eta_{\rm Y}\beta(1-\beta)^2$, and 
 $dW_\beta = \sqrt{2}\real[idV_\Omega e^{-i\theta}]$. 
 
 For any one-dimensional stochastic differential equation of the form (\ref{betaSDE}) on a finite interval  
 there is a simple expression for the stationary probability distribution: 
 \beq
 p\d(\beta) = \frac{N}{B(\beta)}\exp\sq{-\int_\beta^1  \frac{A(\beta')}{B(\beta')}d\beta'},
 \eeq
 where the $N$ is found from normalization \cite{Gar85}. 
In this case it evaluates to 
 \beq
 p\d(\beta) = \frac{N'}{ (1-\beta)^{5/2}}\exp\sq{-\frac{3 \beta(1-\eta_{\rm Y})}{2(1-\beta)\eta_{\rm Y}}}.
 \eeq
Thus the long-time expectation for the square of the conditional average of the Bloch vector length under this monitoring is
\beq \label{avbeta}
{\rm E}[(\an{\s{y}}\d)^2 + (\an{\s{z}}\d)^2] = \int_0^1 p\d(\beta) \beta \, d\beta ,\eeq
which can easily be solved to arbitrary numerical precision. 
 
\subsection*{\large General diffusive unravellings} 

The generalization of the homodyne measurement term 
\beq
d\rho = \sqrt{\gamma}{\cal H}[ dZ^*(t) \s{-}]\rho
\eeq
 to arbitrary diffusive unravellings simply leaves the constraint $\an{ |dZ(t)|^2}= \eta dt$ unchanged, and simply 
 changes  $\an{ dZ^2} = e^{2i\phi}\eta dt$,  to $\an{ dZ^2} = \upsilon dt = e^{2i\phi}|\upsilon|dt$ \cite{WisMil10}. Here $|\upsilon|\in [0,\eta]$  quantifies the extent to which it is possible to obtain information about only one quadrature of the output, 
that defined by the LO phase $\phi$. A given unravelling is thus defined by the pair $(\eta,\upsilon)$, and we use this to label the noise process $dZ_{(\eta,\upsilon)}$. If $\upsilon=0$ there is no selectivity at all (we obtain information equally about all quadratures) and the QSD of Ref.~\cite{GisPer92b} corresponds to using $dZ_{(1,0)}$.  

Now for a given $\eta$ and $|\upsilon|$ it is natural that an experimenter can choose any $\phi$.
If $|\upsilon| \leq 1-\eta$, then for any one of these unravellings $(\eta,\upsilon)$ we can define an independent noise process $dZ_{(1-\eta,-\upsilon)}$ such that
\beq
dZ_{(1,0)} = dZ_{(\eta,\upsilon)} + dZ_{(1-\eta,-\upsilon)}.
\eeq
That is, we could obtain the unravelling $(\eta,\upsilon)$ by starting with the QSD evolution, and {\em discarding} the information in $dZ_{(1,0)}$ relating to the quadrature with phase $\phi + \pi/2$, leaving us only with $dZ_{(\eta,\upsilon)}$. In other words, it is possible that the QSD evolution, driven by the noise $dZ_{(1,0)}$, is the true objective evolution, and all we do by changing $\phi$ is obtain different information about $dZ_{(1,0)}$. To disprove this---to prove that measuring different quadratures genuinely ``steers'' the evolution of the quantum system---thus requires 
$\eta + |\upsilon| > 1$. Since $|\upsilon|\leq \eta$, we  obviously require $\eta>1/2$. 

Note that our proof makes no assumptions about the nature of the quantum system, except that only one of its decoherence channels (e.g. a radiative transition) is monitored. 

\end{widetext}


\begin{thebibliography}{10}

\bibitem{Boh13}
N.~Bohr, Phil. Mag. \textbf{26}, 1 (1913).

\bibitem{Ein17}
A.~{Einstein}, Physikalische Zeitschrift \textbf{18}, 121 (1917).

\bibitem{EPR35}
A.~{Einstein}, B.~{Podolsky}, and N.~{Rosen}, Phys. Rev. \textbf{47}, 777
  (1935).
  
\bibitem{Von32}
J. von Neumann, {\em Mathematical Foundations of Quantum Mechanics}
(Springer, Berlin, 1932);
English translation (Princeton University Press, Princeton, 1955). \blk

\bibitem{Car93b}
H.~J. Carmichael, \emph{An Open Systems Approach to Quantum Optics} (Springer,
  Berlin, 1993).

\bibitem{DalCasMol92}
J.~Dalibard, Y.~Castin, and K.~M\o{}lmer, Phys. Rev. Lett. \textbf{68}, 580
  (1992).

\bibitem{GarParZol92}
C.~W. Gardiner, A.~S. Parkins, and P.~Zoller, Phys. Rev. A \textbf{46}, 4363
  (1992).

\bibitem{Bar93}
A.~Barchielli, Int. J. Theor. Phys. \textbf{32}, 2221 (1993).

%\bibitem{Wis96a}
%H.~M. Wiseman, Quant. Semiclass. Opt. \textbf{8}, 205 (1996).

\bibitem{WisMil10}
H.~M. Wiseman and G.~J. Milburn, \emph{Quantum Measurement and Control}
  (Cambridge University Press, 2010).

\bibitem{KorPRB99}
A.~N. Korotkov, Phys. Rev. B \textbf{60}, 5737 (1999).

\bibitem{Gambetta2008}
J.~Gambetta \ea, Phys. Rev. A \textbf{77}, 012112 (2008).

\bibitem{GisPer92b}
N.~Gisin and I.~C. Percival, J. Phys. A \textbf{25}, 5677 (1992).

%\bibitem{Gis93}
%N.~Gisin \ea, \textbf{40}, 1663 (1993).

\bibitem{WisMil93c}
H.~M. Wiseman and G.~J. Milburn, Phys. Rev. A \textbf{47}, 1652 (1993).

\bibitem{SchPCP35}
E.~Schr\"odinger, Proc. Camb. Phil. Soc. \textbf{31}, 553 (1935).

\bibitem{WisJonDoh07}
H.~M. Wiseman, S.~J. Jones, and A.~C. Doherty, Phys. Rev. Lett. \textbf{98},
  140402 (2007).

\bibitem{CavJonWisRei09}
E.~G. Cavalcanti \ea, Phys. Rev. A \textbf{80}, 032112 (2009).

\bibitem{SJWP10}
D. J. Saunders \ea, %S. J. Jones, H. M. Wiseman, and G. J. Pryde,	
	%ÒExperimental EPR-Steering of Bell-local StatesÓ 
	Nature Physics {\bf 6}, 845 (2010).
	
\bibitem{Smi12}
D. H. Smith \ea, %G. Gillett, M. P. de Almeida, C. Branciard, A. Fedrizzi, T. J. Weinhold, A. Lita, B. 
%Calkins, T. Gerrits, H. M. Wiseman, S. W. Nam, and A. G. White, 
%	ÒConclusive quantum steering with superconducting transition edge sensorsÓ
	Nature Comms. {\bf 3}, 625 (2012).

\bibitem{JonWisDoh07}
S. J. Jones, H. M. Wiseman, and A. C. Doherty,
	Phys. Rev. A {\bf 76}, 052116 (2007).


\bibitem{WisToo99}
H.~M. Wiseman and G.~E. Toombes, Phys. Rev. A \textbf{60}, 2474 (1999).


 \bibitem{epaps} Supplemental Material available at [URL].


\bibitem{MabDoh02}
H.~Mabuchi and A.~C. Doherty, Science \textbf{298}, 1372 (2002).

\bibitem{Ber86}
J.~C. Bergquist \ea, Phys. Rev. Lett. \textbf{57}, 1699 (1986).

\bibitem{Gar85}
C.~W. Gardiner, \emph{Handbook of Stochastic Methods} (Spring\-er, Berlin,
  1985).


\end{thebibliography}
\end{document}